# Coastal Altimetry Products in the Strait of Gibraltar

Jesús Gómez-Enri, Paolo Cipollini, *Senior Member, IEEE*, Marcello Passaro,
Stefano Vignudelli, Begoña Tejedor, and Josep Coca

*Abstract*—This paper analyzes the availability and accuracy of coastal altimetry sea level products in the Strait of Gibraltar. All possible repeats of two sections of the Envisat and AltiKa ground-tracks were used in the eastern and western portions of the strait. For Envisat, along-track sea level anomalies (SLAs) at 18-Hz posting rate were computed using ranges from two sources, namely, the official Sensor Geophysical Data Records (SGDRs) and the outputs of a coastal waveform retracker, the Adaptive Leading Edge Subwaveform (ALES) retracker; in addition, SLAs at 1 Hz were obtained from the Centre for Topographic studies of the Ocean and Hydrosphere (CTOH). For AltiKa, along-track SLA at 40 Hz was also computed both from SGDR and ALES ranges. The sea state bias correction was recomputed for the ALES-retracked Envisat SLA. The quality of these altimeter products was validated using two tide gauges located on the southern coast of Spain. For Envisat, the availability of data close to the coast depends crucially on the strategy followed for data screening. Most of the rejected data were due to the radar instrument operating in a low-precision nonocean mode. We observed an improvement of about 20% in the accuracy of the Envisat SLAs from ALES compared to the standard (SGDR) and the reprocessed CTOH data sets. AltiKa shows higher accuracy, with no significant differences between SGDR and ALES. The use of products from both missions allows longer times series, leading to a better understanding of the hydro-dynamic processes in the study area.

*Index Terms*—Coastal altimetry, data screening, retracking, Strait of Gibraltar (SoG), tide gauge, validation.

## I. Introduction

COASTAL altimetry has become a mature discipline owing to the effort of many research groups and institutions [1].[1]

Manuscript received October 8, 2015; revised February 18, 2016 and April 8, 2016; accepted April 23, 2016. Date of publication May 27, 2016; date of current version August 2, 2016. This work was supported in part by the ALCOVA Project (CTM2012-37839) funded by the Spanish Ministerio de Economía y Competitividad and FEDER, and by the Spanish Ministerio de Educación, Cultura y Deporte under the "Salvador de Madariaga" Programme. The ALES retracker development has been supported in part by the ESA/DUE eSurge (ESA/ESRIN Contract Number 4000103880/11/I-LG).

J. Gómez-Enri, B. Tejedor, and J. Coca are with the University of Cadiz, 11202 Cadiz, Spain (e-mail: jesus.gomez@uca.es; begonia.tejedor@uca.es; josep.coca@uca.es).
P. Cipollini is with the National Oceanography Centre, Southampton SO14 3ZH, U.K. (e-mail: cipo@noc.ac.uk).
M. Passaro is with the Deutsches Geodätisches Forschungsinstitut, Technische Universität München (DGFI-TUM), 80539 Munich, Germany (e-mail: marcello.passaro@tum.de).
S. Vignudelli is with the Institute of Biophysics (CNR), 56124 Pisa, Italy (e-mail: vignudelli@pi.ibf.cnr.it).

[1]Successful initiatives launched in the last decade to improve the retrieval of in-shore altimeter data include PISTACH (*Prototype Innovant de Système de Traitement pour l'Altimétrie Côtière et l'Hydrologie*) funded by CNES (*Centre National d'Études Spatiales*); COASTALT (*Development of Radar Altimetry Data Processing in the Coastal Zone*), eSurge, and CP4O (*Cryosat + Oceans*) supported by the European Space Agency (ESA); and the Spanish-funded ALCOVA (*Coastal Altimetry: Validation of altimeter products near the coast*).

A global analysis of the sea level variability near the coasts using satellite altimeter data is now a realistic prospect by virtue of the availability of new reprocessed data with higher along-track spatial resolutions and better accuracy. However, putting this into effect requires a consistent validation effort.

Reprocessing efforts are targeting the two main factors that compromise the availability and quality of altimeter data near the coasts with respect to open ocean: 1) inaccuracies in the retrieval of geophysical information from the shape of the mean returned waveforms from the reflected surface (this retrieval is normally done by some waveform fitting procedures known as *retracking*) and 2) a poorer characterization of some of the geophysical corrections applied to the data. Present altimetry missions (Cryosat-2, AltiKa, and Jason-2) and near-future ones (Sentinel-3, Jason-3, and Sentinel-6/Jason-CS) minimize the impact of these factors on data quality by virtue of state-of-the-art radiometric performance (Cryosat-2, AltiKa, and Jason-2), use of the Ka-band that allows smaller footprints (AltiKa), and SAR-mode operation (Cryosat-2 and all future missions). For past missions (ERS-1/2, Topex/Poseidon, Envisat, GFO, and Jason-1), more efforts still need to be made in order to include their products in coastal applications and models [2].

A radar altimeter measures the two-way travel time of the emitted/reflected signal/echo and the returned power. The amount of energy received is recorded onboard in a time series called a "waveform." The pulse repetition frequency (PRF) determines the number of waveforms recorded per unit of time. The PRF for Envisat Radar Altimeter 2 (RA-2: one of the instruments used in this work) is 1800 (individual echoes: IEs) per second, i.e., 1800 Hz. The tracker onboard sums incoherently packets of 100 IEs in order to reduce the Rayleigh noise associated with the signals assuming uncorrelated noise between consecutive waveforms [3]. These averaged 18-Hz waveforms are transmitted to ground for postprocessing. The along-track spatial separation between 18-Hz points is about 375 m, but the corresponding footprint has a diameter varying from ∼1.6 to 10 km depending on sea state [4]. The *retracking* of waveforms over the ocean is made, assuming the Brown waveform model [5], [6], and yields three parameters: epoch ($t_0$), which is used to estimate the satellite's distance to the mean reflected surface (retracked *Range*), the amplitude of the received signal: backscatter coefficient (*sigma0*) related to the wind speed at the sea surface ($U_{10}$), and significant wave height (SWH). Inaccuracies in the estimates of the retracked *Range* near the coasts are mainly due to the contamination of the waveforms [7]. This contamination might be due to the proximity of land [8] or patches of calm water [9], [10]. In any cases, the effect over the waveform is often clearly seen in the trailing and leading edges.

The way in which this contamination affects the *retracking* of the contaminated waveforms, and hence the accuracy of the aforementioned parameters, is still a matter of investigation. Different strategies have been proposed to mitigate these effects. They are summarized in [11]. Among the various retrackers proposed, we consider the Adaptive Leading Edge Subwaveform (ALES hereinafter) as this has been validated for both *Range* and SWH for different missions (Jason-1, Jason-2, and Envisat) in a few locations [11]–[13]. ALES belongs to the family of retrackers restricting the fitting only to that part of the waveform containing most of the oceanographic information, i.e., the leading edge [14]–[18]. The tail of the waveform, more prone to contamination by bright targets in the footprint area, is not considered in the fitting process. ALES, in particular, is a two-pass retracker: the first pass is focused on the leading edge and gives an initial estimate of the SWH; this value is then used to optimize the width of the subwaveform retracked in the second pass. The ALES algorithm is described in [11], and in the same study, ALES-derived sea level was validated against tide gauges in Trieste (Northern Adriatic-Italy; for Jason-1/2 and Envisat) and Mossel Bay (South African coast; for Jason-2 and Envisat). Validation showed clear improvements in terms of both quality and quantity of recovered data w.r.t. levels in the Sensor Geophysical Data Record (SGDR) products, which are based on a conventional Brown-model retracker [5]. ALES has also been validated for SWH in the German Bay [12], demonstrating that ALES is also able to increase the precision of the SWH estimations compared to the SGDR products, and more recently, ALES sea level has been successfully compared with data from the ESA sea level Climate Change Initiative and from tide gauges in the Danish Straits to assess the sea level annual cycle with a view to climatic applications [13].

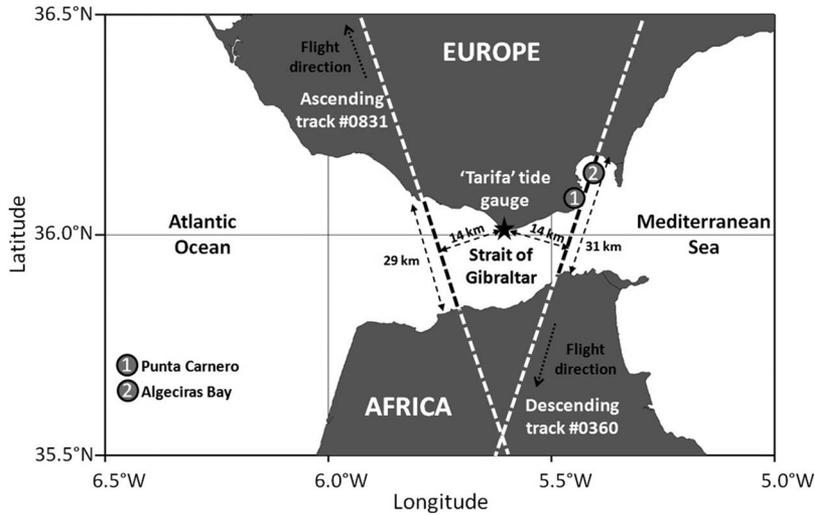

Fig. 1. Study area: the SoG located between Africa and Europe. Also included are the position of the tide gauges and the location of the two passes analyzed: Ascending pass #0831 and descending pass #0360. The length of the "ocean" track segments used and the distance to the tide gauge are also included.

In this work, we analyze in detail the availability and accuracy of altimeter-derived sea level data from Envisat RA-2 and AltiKa SARAL (Satellite with ARgos and ALtiKa) in the area of the Strait of Gibraltar (SoG). Here, Envisat and AltiKa have one 35-day repetitive descending pass in the eastern side of the strait and one ascending in the western side. We assess the accuracy of sea level altimeter data using the time series of two tide gauges located in the Spanish coast between both passes.

We analyze the performance of ALES in comparison with the official SGDR product based on [5]. To do this, we estimate the relative root-mean-square error (rmse) between concomitant altimeter and tide gauge data in a few land/ocean transition scenarios along the eastern and western sides of the strait. Section II of this paper presents the study area. The data sets used (altimeter, tide gauge, and auxiliary data) are illustrated in Section III. Section IV describes the methodology adopted to create the time series of sea level anomaly (SLA) from the altimeter and the tide gauge. Section V presents the results both in terms of analysis of the availability of altimeter data and in terms of their accuracy, i.e., along-track rmse between the altimeter and tide gauge time series. These results are discussed in Section VI, and the conclusion is given in Section VII.

## II. STUDY AREA

The SoG is between the Iberian Peninsula and northern Africa: [$35.75°$–$36.20°$ N]–[$-5.90°$ W–$-5.25°$ W] (Fig. 1). It is the unique connection between the Atlantic Ocean and the Mediterranean Sea and controls the water exchanges between both water masses. The Algeciras Bay (*alg-Bay*) is located near latitude $36.2°$ N, at the northeastern end of the strait. The SoG has been thoroughly described in the past from different points of view. References [19] and [20] analyzed the surface flux of Atlantic water toward the East being compensated by a western flux of Mediterranean deeper, saltier, and warmer water. The seasonal and interannual oscillations of these fluxes [21]–[24] (among others) are responsible for a sea level difference observed between the Atlantic and the Mediterranean Sea that might be driven by different forcing mechanisms: tides [25], atmospheric pressure variations [26], steric contributions [27], geostrophic controls inside the strait [28], and winds in the surrounding area [24], [29], [30]. In addition to this quasi-steady two-layer water exchange, a mesotidal and semidiurnal tide dynamics is observed [31]–[35]. The water flow interaction with the topography (Camarinal Sill) in the western side of the strait under certain hydrographic conditions generates a train of internal waves, which move mainly toward the Mediterranean Sea [36]–[40].

From an altimetric point of view, [29] and [41] analyzed the sea level difference between the Atlantic Ocean and the Mediterranean Sea near the strait using Topex/Poseidon tracks. However, they only used along-track altimeter data at 1-Hz interval (about 6 km along the ground track) in regions deeper than 1000 m at distances greater than 150 km from the eastern and western sides of the strait. They pointed out the lack of accurate altimeter data for shallower regions. More recently, [42] developed a preliminary analysis on Envisat altimeter data availability and accuracy in the study area.

## III. DATA SETS

Two passes of Envisat/AltiKa were available in the study area: a descending and an ascending crossing the eastern/western side of the strait, respectively. These are the only satellites with two repetitive passes inside the limits of the SoG. The presence and orientation of these tracks in the SoG and their relative vicinity to the tide gauges offer a good opportunity to test the quality of coastal altimetry measurements in different land-to-ocean and ocean-to-land transitions. The minimum distance between the satellite's passes (ascending and descending) and the tide gauges was about 14 km (Fig. 1). We defined three along-track segments of interest: Algeciras Bay (*alg-Bay*: 11.0 km long) and Eastern SoG (*E*-SoG: 18.0 km) for the descending pass (#0360) and Western SoG (*W*-SoG: 29.0 km) for the ascending (#0831). A high-quality altimeter-derived coastal product over those two passes would allow some degree of continuity from the two missions (except, of course, for the 2.5-year gap between the end of the Envisat phase E2 and the start of AltiKa measurements, as detailed in the following), leading to a better understanding of the hydrodynamic processes at both sides of the strait, which is the ultimate motivation for the present assessment study.

### A. Envisat RA-2

ESA's satellite Envisat was launched in March 2002, being in operation for about 10 years. The satellite had a sun-synchronous quasi-polar orbit with a 35-day repeat cycle (phase E2) that changed to a 30-day orbit in October 2010 until the end of the mission in April 2012 (phase E3). In this work, we focused on the first longer 35-day repeat cycle. The time period analyzed spanned 8 years from October 2002 (cycle 6) to October 2010 (cycle 93), giving a maximum of 88 cycles. The passes of Envisat RA-2 available in the study area were the following: descending #0360 (D#0360) crossing the study area at about 10:46 UTC time in the eastern side of the strait and ascending #0831 (A#0831) crossing at about 21:58 UTC time in the western side (Fig. 1).

*SGDR:* In this work, we used 18-Hz data from the latest official SGDR product under Version 2.1 (which accounts for satellite orbit evolution and implements the Ultra Stable Oscillator instrumental correction). The information extracted from the SGDR files was the following: coordinates (time and measurement position; 18-Hz posting rate), *Orbit altitude* (18 Hz), *Range* (ocean retracker at Ku-band based on [5]; 18 Hz), "range" corrections (1 Hz, linearly interpolated to 18 Hz), "geophysical" corrections (1 Hz, linearly interpolated to 18 Hz), and the Ku-band waveforms (18 Hz). SLA along the two track segments analyzed was obtained as detailed in the next section.

*ALES:* Along-track retracked *Range*, SWH, and *sigma0* from the ALES retracker were used to estimate SLA at 18-Hz posting rate. We retracked the waveforms of the two track segments available in the SGDR product in the study area along the analyzed time period.

*CTOH:* Data from the Centre for Topographic studies of the Ocean and Hydrosphere (CTOH; http://ctoh.legos.obs-mip.fr/products/alongtrack-data/alongtrack-data/) were obtained from the X-TRACK processor and were distributed by Aviso (Archiving, Validation, and Interpretation of Satellite Oceanographic). This is a Level 3 product with data availability at approximately every 7 km along the Envisat passes analyzed (1-Hz posting rate). X-TRACK does not retrack the waveforms but aims at improving the availability and accuracy of sea level measurements in coastal zones through more accurate tidal and atmosphere forcing corrections, data editing, and filtering [43], [44].

### B. AltiKa SARAL

AltiKa is a cooperative mission between the Indian Space Research Organisation and the French National Centre of Space Research (CNES). The descending/ascending passes of AltiKa's SARAL altimeter cross the study area at about 18:51/06:02 UTC time, respectively (Fig. 1). Sea level data at 40-Hz posting rate were obtained from the official SGDR product available at the Aviso ftp server: avisoftp.cnes.fr/Niveau0/AVISO/pub/saral/sgdr_t/. The time period was May 2013 to January 2015 (19 cycles). The retracked *Range* available from the SGDR is estimated by a maximum likelihood estimation approach: MLE3 full-waveform fitting algorithm that uses the Brown analytical model [5]. The ALES retracker was also applied to the waveforms to estimate the *Range*.

### C. In Situ Data

Two tide gauges were used for comparison against altimeter data: Tarifa_ENV and Tarifa_ALT for Envisat/AltiKa, respectively.

*Tarifa_ENV:* The tide gauge was located in the harbor of Tarifa city: [$36.0086°$ N–$-5.6026°$ W] being in operation from 1943 to 2012 (Fig. 1). It recorded water levels at 5-min interval referred to the tide gauge zero (TGZ) with no activity during two years (1962 and 1990) and in some other sporadic periods of time. It was part of the Spanish Institute of Oceanography (IEO) Network and fulfilled the Global and European Sea Level Observing Systems requirements (GLOSS and EOSS, respectively) [45], [46]. It was part of the Permanent Service for Mean Sea Level (PSMSL) network (http://www.psmsl.org). The measurement system was composed of two instruments: a mechanical float tide gauge and an electromagnetic codifier (Allgomatic data logger) for converting the lineal movement of the wire float to a digital value, with millimeter precision [47].

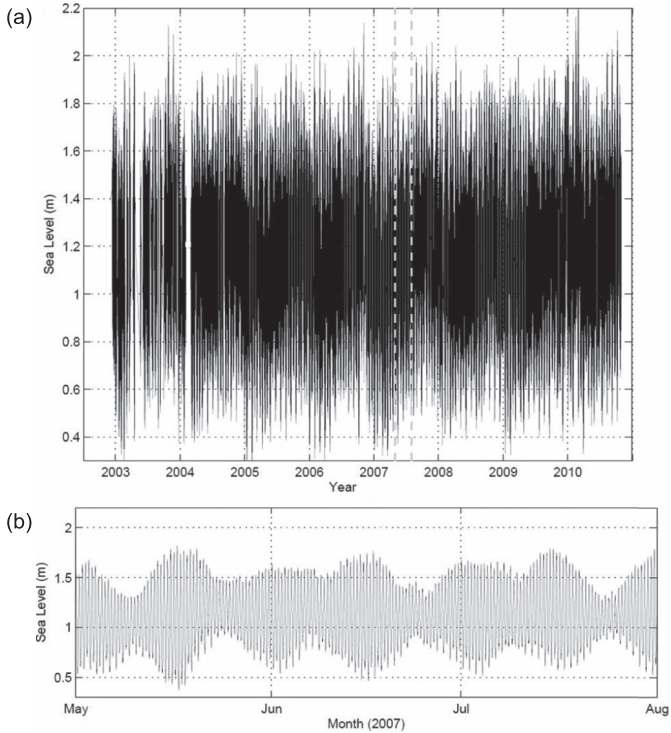

Fig. 2. Sea level (in meters) recorded by tide gauge: *Tarifa_ENV* at 5-min interval during the Envisat time period (October 2002 to October 2010). Data are referred to the TGZ. (b) Zoom-in of the water level between May and July 2007.

Fig. 2 shows the instantaneous 5-min water level recorded by the tide gauge [Fig. 2(a)] along the time period of comparison against Envisat data (a lack of data was observed between October and December 2002 and in February 2004). A zoom-in between May and July 2007 [Fig. 2(b)] clearly shows the semidiurnal tides dominating the signal. The monthly average of the water level (not shown) indicates a clear seasonal cycle in most of the years with an interannual variability.

*Tarifa_ALT:* The instrument is a MIROS (MIcrowave Remote sensor for the Ocean Surface) radar sensor measuring at 2 Hz located at approximately the same position as Tarifa_ENV: [36.0065° N–−5.6035° W]. Data are then averaged to 1-min intervals at the instrument before transmission in real time to a processing facility where a final 5-min product is generated for distribution. Data are available for the period from July 2009 to the present. The tide gauge is managed by the Spanish Puertos del Estado (http://www.puertos.es) and belongs to the Red de Mareógrafos (REDMAR) network of Puertos del Estado. REDMAR is integrated in the PSMSL and GLOSS.

### D. Auxiliary Data

Some of the corrections used to create time series of SLA were applied to both data sources, altimeter and tide gauge: tidal elevation and atmospheric effect.

*Tidal Model:* We used the National Space Institute of Danmarks Tekniske Universitet (DTU) DTU10 global ocean tide model [48]. This is an updated version of the AG95 (Andersen-Grenoble) ocean tide model with a resolution of 0.125° × 0.125° based on the finite element solution, FES2004 [49].

We used the routines provided by DTU to estimate the total geocentric tidal elevation for the time and position of every 18/40-Hz Envisat/AltiKa data point along the two tracks. The same routines were used to detide the water level from the tide gauges.

The performance of this model at the Tarifa_ENV location was checked. We applied a harmonic analysis to one year (2009) of tide gauge data and obtained the main constituents. We then estimated the *in situ* ocean tide (local tide) at the time of the Envisat data at the tide gauge location. The percentage of explained variance by the DTU10 and local tide was calculated as follows [50], [51]:

$$\% \text{ var} = 100 \left(1 - \frac{\sigma^2_{\text{residual}}}{\sigma^2_{\text{original}}}\right) \quad (1)$$

where $\sigma$ stands for the standard deviation of the time series, *original* refers to the uncorrected sea level, and *residual* refers to the detided time series using DTU10 and local tide, respectively. By applying (1), we found that DTU10/local tide explain 93%/95% of the sea level variance in both tracks. We also estimated the root-mean-square (rms) misfit between the main constituents derived from the tide gauge and the constituents provided by DTU10 as in [52]. The rms of the constituents ($M_2$, $S_2$, $N_2$, $K_2$, $K_1$, $M_4$, $O_1$, $P_1$, and $Q_1$) is below 4 cm in all cases, with a root-square sum of 4.6 cm. Thus, DTU10 seems to accurately model the tides in the study area.

*DAC:* The altimeter data use a dynamic atmospheric correction (DAC) to correct for the effects of high-frequency winds and atmospheric pressure oscillations with periods lower than 20 days and the inverted barometer correction [53]. *DAC* is computed with the high-resolution 2-D barotropic model MOG2D ("Modèle aux Ondes de Gravité"). The lack of information regarding the winds precluded the estimation of its contributions to the *in situ* water level. For this reason, we used the regular 6-hourly gridded maps of *DAC* from AVISO to correct these atmospheric effects to the data sets. The correction was estimated, interpolating the *DAC* maps to the time series and positions of altimeter and tide gauge data sets.

*MSS:* We used the most updated version of the DTU mean sea surface (MSS): DTU13 [54], [55]. The spatial resolution is 1 min by 1 min. DTU13 was interpolated to the time series of along-track positions of the two tracks of Envisat and AltiKa.

### IV. METHODOLOGY

From the time series of the tide gauges, we extracted the water levels at the two closest times to each altimeter measurement. Tide gauge and altimeter data sets were collocated in time using the satellite measurement as reference interpolating the *in situ* water level to the exact time of the radar records. We analyzed the availability of concomitant *in situ* and altimeter data. After the collocation, we obtained a maximum of 66/18 (Envisat/AltiKa) pairs of *in situ* and altimeter data along track in D#0360 and 74/18 cycles in A#0831. The discrepancy in the number of collocated data in Envisat with respect to the maximum number of cycles (88) was due, first, to the unavailability of *in situ* data in some of the dates of the radar measurements and, second, to the lack of some altimeter cycles. We computed time series of SLA from altimetry data: Envisat (SGDR–ALES: 18-Hz posting rate, and CTOH: 1 Hz) and AltiKa (SGDR–ALES: 40 Hz). The concomitant time series from the tide gauges were obtained following the posting rates of the altimeter products used. The range and geophysical corrections used from the Envisat SGDR files are provided at 1 Hz, so they were linearly interpolated to 18 Hz. In the case of AltiKa, these corrections were available at 40 Hz.

### A. SLA From Altimetry

The SLA was obtained following (2):

$$\text{SLA} = \text{Orbit} - \text{Range} - \text{Range Corrections} - \text{Geophysical Corrections} - \text{MSS}. \quad (2)$$

*Orbit:* Is the distance between the satellite's orbit and a reference surface: ellipsoid WGS84 for Envisat and the ellipsoid used by the Topex-Poseidon, Jason-1, and Jason-2 missions for AltiKa.

*Range:* The retracked *Ranges* used in this work for Envisat/AltiKa were the following: 1) from the ocean retrackers at Ku-/Ka-bands available in the SGDR products [5] and 2) from the ALES retracker [11].

*Range Corrections:* The ionospheric correction applied to Envisat/AltiKa data sets was the global ionospheric maps based on total electron content grids developed by the Jet Propulsion Laboratory. The dry/wet tropospheric corrections applied to both missions were obtained from the European Centre for Medium Weather Forecast model computed by Météo-France, the French Meteorological Agency.

The Sea State Bias correction (SSB_SGDR_Env) applied to Envisat_SGDR was obtained by bilinear interpolations from a look-up table which is a function of SWH and $U_{10}$ derived from one year of Envisat RA-2 Ku-band waveform retracking [56]. For AltiKa (SSB_SGDR_Alt), the same methodology is applied from one year of data. In addition, SWH and *sigma0* obtained from Envisat_ALES were used to recompute the *SSB* correction (SSB_ALES_Env, hereinafter) for the retracked *Range*. To do this, *sigma0* was converted to $U_{10}$ by using the algorithm described in [57]. Basically, the algorithm uses a first-guess estimation of $U_{10}$ ($U_m$) obtained by fitting a two-segment function (one linear and one exponential) to *sigma0*. SSB_ALES_Env was then estimated by bilinear interpolation from the look-up table in [56] using SWH and $U_{10}$ from ALES as inputs. Note that, as mentioned previously, SSB_SGDR_Env is interpolated to 18 Hz from the 1-Hz averages; conversely, SSB_ALES_Env is computed natively at the higher rate, so its 18-Hz samples will show high-frequency variability.

*Geophysical Corrections:* As mentioned, the tidal elevation used was the DTU10 tidal model for both Envisat and AltiKa. Solid Earth Tide and Pole Tide were also added from SGDR. The atmospheric effects were removed by the interpolated DAC.

Four time series (at 18 Hz) were obtained for Envisat in the two track segments analyzed: 1) SLA_Envisat_SGDR_{D#0360; A#831} with the *Range* and SSB_SGDR_Env coming from the SGDR files based on the Ocean retracker and 2) SLA_Envisat_ALES_{D#0360; A#831} with *Range* and SSB_ALES_Env obtained from the retracking of the waveforms using the ALES retracker. Two time series (at 1 Hz) were obtained for Envisat CTOH: SLA_Envisat_CTOH_{D#0360; A#831}. Finally, four time series (at 40 Hz) for AltiKa: SLA_AltiKa_SGDR_{D#0360; A#831} and SLA_AltiKa_ALES_{D#0360; A#831}.

A measure of the improvement due to the SSB_ALES_Env correction (Envisat data) is the reduction in the uncertainty of the sea level on the two track segments crossing the strait, which we computed as in [58] using the outputs of the ALES retracker. The uncertainty drops from 22.1 and 16.6 cm to 20.8 and 14.2 for D#0360/A#0831, respectively, when SSB_ALES_Env is applied to the SLA_Envisat_ALES instead of the SSB_SGDR_Env.

### B. SLA From Tide Gauges

With the *in situ* time series of water levels interpolated to the exact time of the altimeter measurements of the two passes analyzed (Envisat and AltiKa), we obtained the SLA as

$$\text{SLA} = \text{Water\_Level} - \text{Geocentric Ocean Tide} - \text{DAC} - \text{MSS}. \quad (3)$$

- *Water_Level* is the record interpolated to the time of the altimeter measurement.
- *Geocentric Ocean Tide* was extracted from the DTU10 global ocean tide model using the location of the tide gauge and the time of the altimeter data as references.
- The atmospheric effects were removed using the interpolated DAC.
- *MSS*: is the mean sea level (1990–1999) over the TGZ.

The *in situ* time series were the following: SLA_TG_Envisat_18Hz_{D#0360; A#831} and SLA_TG_Envisat_1Hz_{D#0360; A#831} for comparison against Envisat (18-Hz and 1-Hz products, respectively); and SLA_TG_AltiKa_40Hz_{D#0360; A#831} for AltiKa.

### C. RMSE

The quality of the altimeter SLA time series was made by estimating the relative rmse between the time series of Envisat/AltiKa from both retrackers and the equivalent time series of the tide gauges. This parameter (also known as rms difference) has been thoroughly used to estimate the validity of coastal altimeter data [11], [59], [60] (and the references therein). We performed a relative analysis as no information on the ellipsoidal height of the tide gauges was available. The relative rmse was computed by removing the temporal mean of the time series before comparison.

## V. RESULTS

The results of this study are of two kinds. First is the results in terms of data availability (i.e., data quantity) with an analysis of what causes the data dropouts. This is particularly important for Envisat which has chirp bandwidth issues as discussed in

Section V-A. Then, there are results from the validation against tide gauges, allowing a quantification of the accuracy (i.e., data quality) for oceanographic applications; these are presented in Section V-B.

### A. Availability of the Coastal Altimetry Records

Here, we analyze the factors affecting the screening out of altimeter data which is necessary before performing the comparison against *in situ* data. Three conditions were taken into account for altimeter data rejection: 1) for Envisat RA-2 only, the instrument can be operating in a low-precision nonocean chirp bandwidth; 2) the bad quality of the corrections; and 3) The presence of SLA outliers.

*RA-2 Chirp Bandwidth:* Envisat RA-2 was designed to operate at three different chirp bandwidths in Ku-band, depending on the type of surface: 320 MHz (corresponding to a pulse length of 3.125 ns, i.e., a resolution of ∼47 cm for the single pulse) for ocean zones and 80 or 20 MHz for nonocean surfaces. In open ocean conditions, the waveform shapes have smooth variations over a few seconds; hence, RA-2 could use the highest resolution without losing tracking of the surface. Over rapidly changing topography (i.e., coastal zones) where the tracking could be lost, the instrument operated in coarser resolutions, preventing the interruption of the echo sample collection. The Brown and ALES retrackers used in this work have so far only been implemented for the ocean-type (320 MHz) waveforms. Thus, only the radar measurements (*Range*) obtained by retracking waveforms with a chirp bandwidth of 320 MHz were taken into account. Measurements taken with lower bandwidths (80 and 20 MHz) have intrinsically much lower precision and resolution (by a factor of 4 and 16, respectively), thus making their use not recommended anyway. Fig. 3 presents two examples of radargrams showing the waveform shapes (power) along the two track segments analyzed: D#0360 [Fig. 3(a)] and A#0831 [Fig. 3(c)]. For the examples, we chose orbital cycle number 73 in both cases as the passes in this cycle show all of the factors affecting the loss of data identified. We included the corresponding SLA_ALES profiles [Fig. 3(b) and (d) for descending and ascending passes, respectively]. The unavailability of radar measurements due to the instrument operating in a nonocean mode is observed in the northern land-to-ocean transition of D#0360 (20% of waveforms) and the southern transition of A#0831 (30%). In the southern-D#0360/northern-A#0831 ocean-to-land transitions, there is no loss of data as the instrument was operating in ocean mode very close to the land. The radar instrument rapidly changed its chirp bandwidth from 80 to 320 MHz and then back to 80 MHz in this specific cycle in *alg-Bay* [Fig. 3(a)]. The small width of the bay (∼7 km) complicates the interpretation of the "ocean" waveform shapes due to land contamination in the footprint area.

Fig. 4 summarizes the RA-2 data availability considering all of the cycles. It shows the number of cycles along the two track segments analyzed having a chirp bandwidth of 320 MHz (black solid line). The number of cycles in "ocean" mode increases steadily for D#0360 [Fig. 4(a)] inside the bay (*alg-Bay*) from the northern land-to-ocean transition to Punta Carnero.

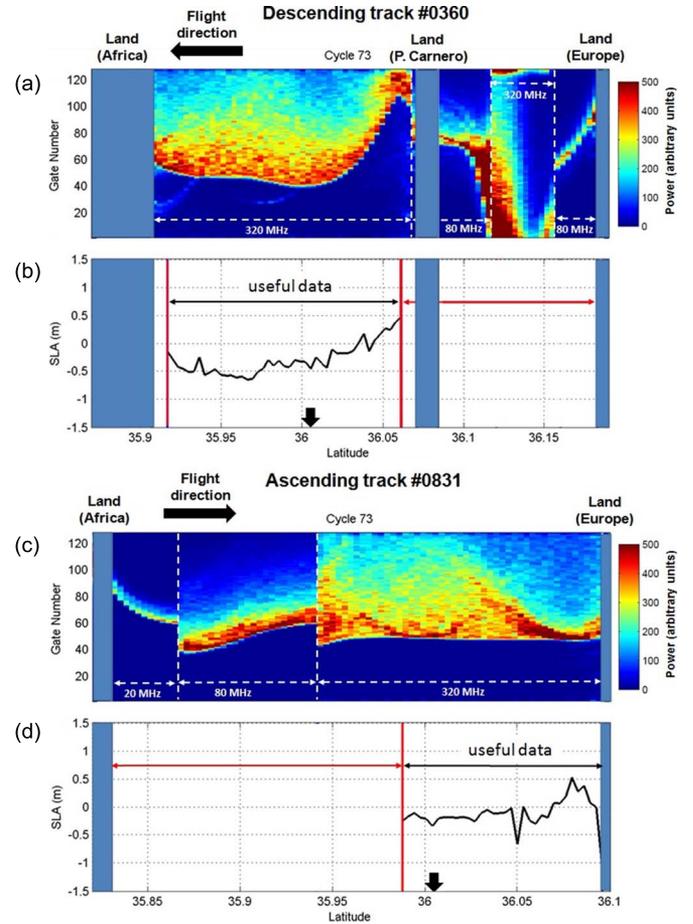

Fig. 3. Envisat RA-2 radargrams of along-track waveform power for descending D#0360 (a) and ascending A#0831 (c) track segments with the chirp bandwidth also included. The selected cycle was number 73 in both cases. The along-track SLA profiles (useful data) are shown in (b) (D#0360) and (d) (A#0831). Red arrow indicates the segments with rejected data after screening. The big black arrows give the latitudinal position of the tide gauge.

Most of the cycles are in this mode in the strait (*E*-SoG) even in the southern part of the track when the satellite approaches its ocean-to-land transition. In ascending track (A#0831), we observe a low number of cycles in ocean mode in the first 10.5 km of the track segment [Fig. 4(b)] in the southern land-to-ocean transition. Then, the data availability increases steadily in the second sector of the track (of about 10 km long). Finally, the percentage is almost 100% in a third sector (8.5 km) in the northern track segment.

These results in the SoG confirm that, for Envisat RA-2, the availability of data in ocean mode (320 MHz) depends significantly on the type of land/ocean transition. In ocean-to-land transitions, we observe, on average, a higher number of "ocean" waveforms than in land-to-ocean transitions. The complex topography of the land makes the radar operate in coarser resolutions in land-to-ocean transitions, and it takes some time to switch back to ocean mode. As said, in the remainder of our analysis, we only consider data acquired in ocean mode.

*Along-Track Availability of "Range"/"Geophysical" Corrections:* For any altimeter, we expect some loss of data due to poor accuracy of some of the range and geophysical corrections applied to estimate SLA in the vicinity of land, resulting in SLA outliers. To quantify this issue, we determined the number of along-track cycles with corrections inside their validity range. All of the corrections used from the SGDR files showed 100% of availability for both passes and both missions. The only exception to this was the SSB_SGDR_Env/SSB_ALES_Env for Envisat and SSB_SGDR_Alt for AltiKa. This correction is obtained using information gathered from the retracking of the waveforms (SWH and $U_{10}$) [56]. The retrieval of these parameters in the coastal zone might be affected by land reflections in the footprint area. This would, in turn, lead to inaccurate estimates of SSB. Taking into account the transitions observed in the track segments analyzed, we might expect a number of data rejections due to invalid SSB for both missions. For SSB_SGDR_Env/SSB_SGDR_Alt, "invalid" means the values outside the expected range of variation for SSB: $[-0.5 - 0]$ m [61]. For SSB_ALES_Env, invalid values were those obtained with SWH and $U_{10}$ input values (from ALES retracker) bigger than the upper limits of the look-up table used (12 m and 20.75 m/s, respectively; the number of invalid values might be reduced with some degree of along-track smoothing of the native 18-Hz SSB_ALES_Env, which is the scope for future work). For Envisat, the impact of the screening based on the corrections on top of the chirp-based one is shown by the green solid lines in Fig. 4. More cycles were lost due to invalid SSB_SGDR_Env/SSB_ALES_Env values in *alg-Bay* and at the ocean-to-land/land-to-ocean transitions in the strait. The number of valid cycles increases as the satellite approaches open ocean conditions. For AltiKa, the screening based on the corrections (not shown) confirms the loss of cycles due to invalid SSB_SGDR_Alt.

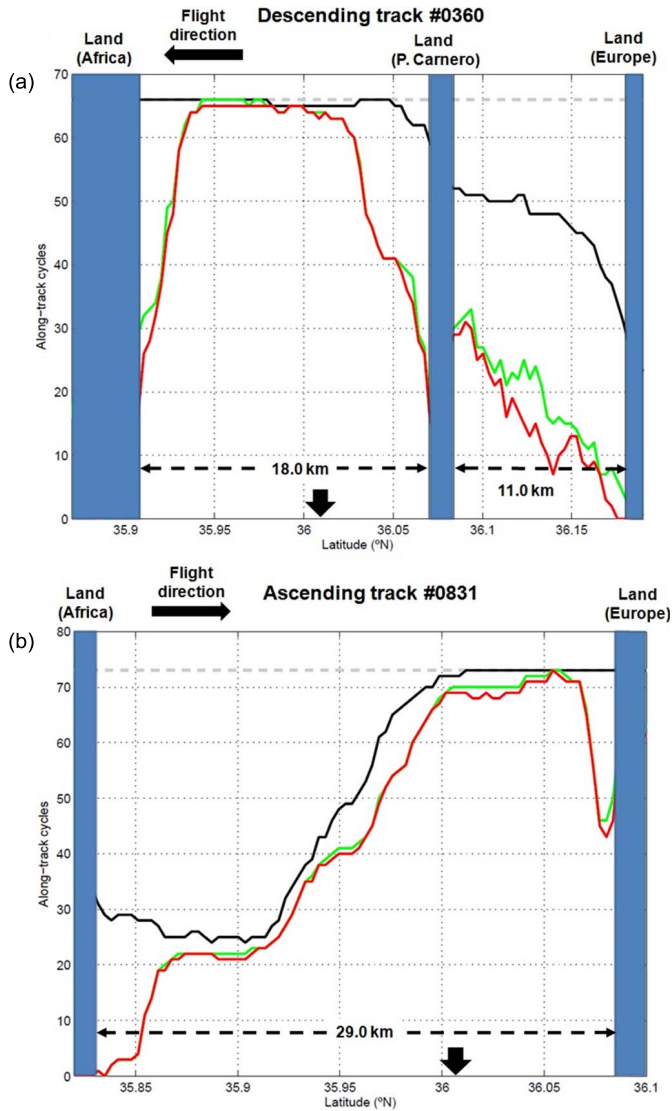

Fig. 4. Envisat data availability (# of cycles) along the two tracks analyzed: D#0360 (a) and A#0831 (b). The gray dashed line gives the maximum number of cycles: 66 (D#0360) and 74 (A#0831). The black solid line indicates the number of cycles after applying the chirp_id mask to the data set. The green solid line gives the number of along-track cycles used to estimate SLA after applying the editing of the corrections. The red solid line shows the number of cycles after all of the outliers in SLA were removed. The big black arrows give the latitudinal position of the tide gauge.

*Removal of SLA Outliers:* Taking into account only "ocean" radar measurements and corrections within their range of validity, we estimated the time series of SLA (2) along the two tracks for both missions. We considered only SLA values within $[-1.5\ 1.5]$ m. This gave the final number of cycles for comparison against *in situ* SLA. The Envisat cycle analyzed in Fig. 3 (73) shows the track segments rejected due to the following: 1) areas where the chirp bandwidth was not 320 MHz; 2) invalid SSB; and 3) SLA out of its range of validity [delimited by red arrows in Fig. 3(b) and (d)]. The lack of these data is observed in *alg-Bay* and close to the southern ocean-to-land transition in D#0360 [Fig. 3(b)]. For A#0831 [Fig. 3(d)], some data rejection is observed following the land-to-ocean transition and extends to the first measurements made by the instrument operating in ocean mode.

Fig. 4 shows the impact of SLA outlier screening for Envisat as red solid lines. For the sake of comparison among retrackers, we only considered altimeter time series in locations in which both SGDR-derived and ALES-derived SLAs were available after the screening. A few more cycles are lost in most of *alg-Bay* [Fig. 4(a)]. The availability of valid data continues to increase in *E*-SoG to reach almost its maximum (66). A small dropout is observed due to the proximity of land as the satellite approaches the southern ocean-to-land transition. Regarding ascending A#0831 [Fig. 4(b)], the loss of data due to data screening is only observed in a few locations.

It is interesting to discuss what causes the rejection of so many records in *alg-Bay* (D#0360). The altimeter is in the correct ocean bandwidth mode in more than half of the passes, as in the last few kilometers flown over land before the coastline where the terrain has only moderate slope. However, there are difficulties with the corrections especially the SSB as discussed in Section V-A2, which result in the rejection of many records. A few more outliers remain in the SLA in the center of the bay, likely to be a result of the several "bright targets" (calm water in sheltered areas; see [9]) surrounding it. The corresponding land-to-ocean transition of track A#0831 has a higher proportion (55% to 65% in the first 8 km from the coast) of nonocean bandwidth records due to the more corrugated terrain over the African coast (the track overflights a 450 m relief at 8 km from the coastline) but a much smaller proportion of rejections due to corrections or SLA outliers.

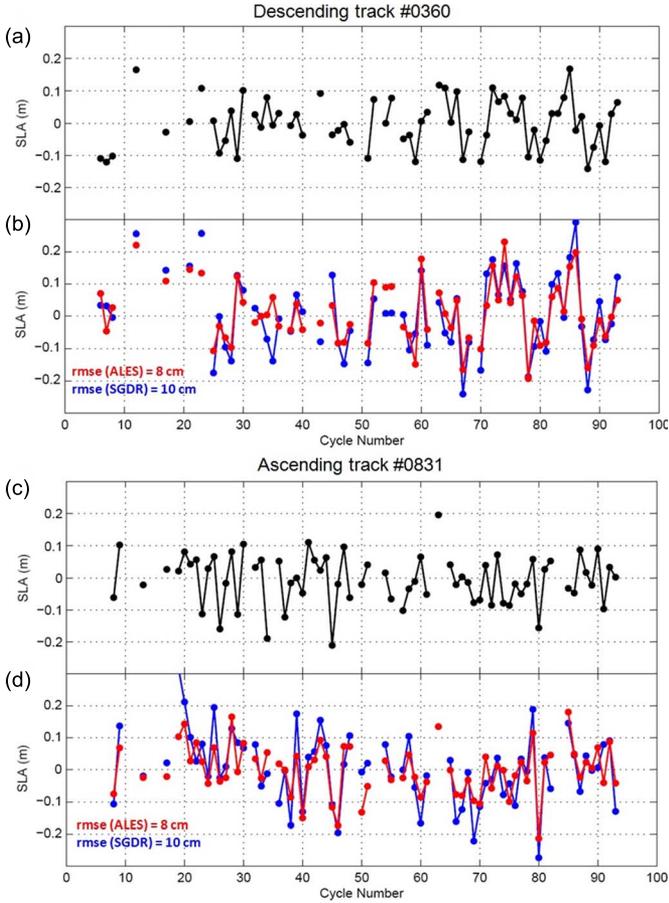

Fig. 5. Time series of *in situ* SLA (a) and (c) and altimeter-derived SLA: SGDR (blue line) and ALES (red line) (b) and (d) for descending pass #0360 and ascending pass #0831, respectively. The 18-Hz position selected was at the lowest rmse found at both ALES and SGDR data sets along the entire track segments.

*B. Validation of Altimeter-Derived SLA*

The altimeter data editing generates times series of SLA along the two tracks analyzed. These along-track time series were compared with the concomitant time series of SLA obtained from the tide gauges. Fig. 5 shows the time series of SLA tide gauge: SLA_TG_Envisat_18Hz_D#0360 [Fig. 5(a)], Envisat descending pass: SLA_Envisat_{SGDR; ALES}_D#0360 [Fig. 5(b)], SLA_TG_Envisat_18Hz_A#0831 [Fig. 5(c)], and ascending pass: SLA_Envisat_{SGDR; ALES}_A#0831 [Fig. 5(d)]. We selected the 18-Hz position with the lowest rmse. The distance to the nearest tide gauge was about 15 km in both along-track points. The lack of data is mainly observed at the beginning of the time period selected. Tide gauge SLA series ranges between −0.2 and 0.2 m, with most of the altimeter SLA values (SGDR and ALES) inside that range. The rmse between *in situ* and altimeter time series in the along-track points selected was 8/10 cm (ALES/SGDR for each track segment).

Fig. 6 shows the rmse obtained along the two tracks analyzed. We only plotted the results in the along-track positions with at least 20 valid RA-2 cycles. We included the comparison made using the Envisat SLA obtained from CTOH. In general, the along-track rmse in #D0360 [Fig. 6(a)] ranges between

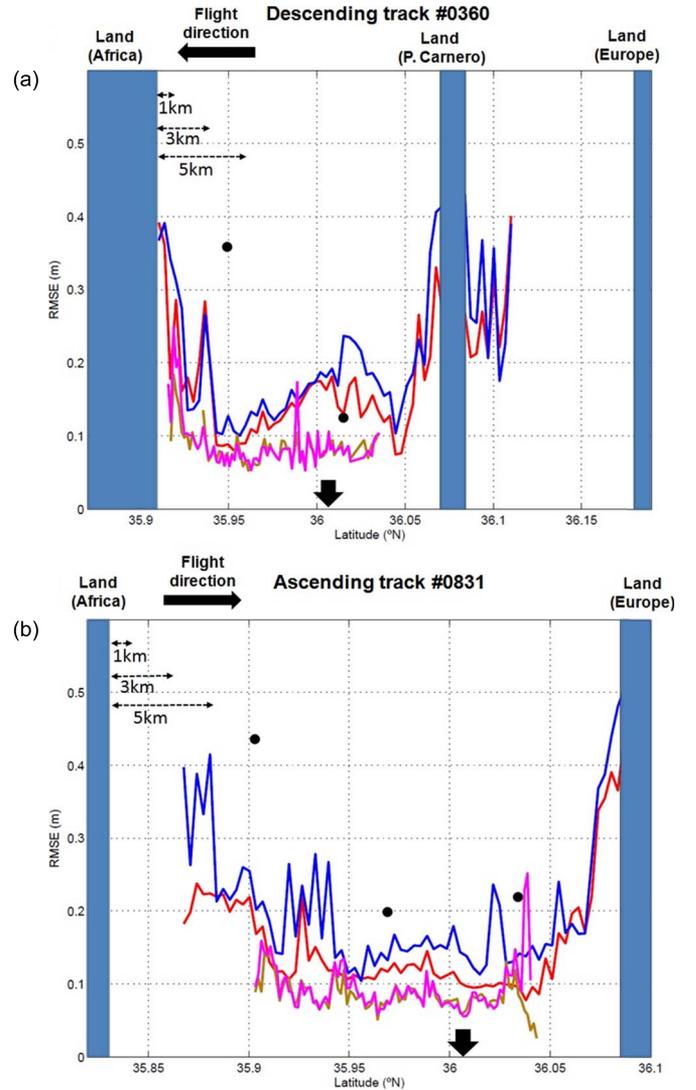

Fig. 6. RMSE along the two track segments analyzed: D#0360 (a) and A#0831 (b). The blue lines show the results obtained with Envisat SGDR, and the red lines show those from Envisat ALES. Black dots are the rmse for CTOH Envisat data set (1 Hz). Also included the rmse from AltiKa/standard (pink line) and AltiKa/ALES (brown line).

8 and 40 cm (ALES/SGDR), with the higher values observed in land/ocean transitions (lower number of available cycles). The lower rmse is observed at ∼14/15 km from the TG location. In this particular track, ALES seems to perform better than SGDR in most of the segment. Only two 1-Hz CTOH points were obtained in this track, showing similar rmse to ALES for the closest CTOH point to the tide gauge position. The rmse for #A0831 [Fig. 6(b)] ranges between 8 and 50 cm. We observe decreasing values as the track crosses the strait northward. *RMSE* is higher in the land/ocean transition. Over this track, ALES performs much better than SGDR in terms of lower rmse. Only three 1-Hz CTOH points were obtained for this track with rmse values higher than both ALES and SGDR. The improvement of ALES with respect to SGDR confirms previous analysis made in [11].

We included in Fig. 6 also the along-track rmse using AltiKa, which was screened as per Envisat (except, of course, for the chirp issue). AltiKa presents a lower rmse (below 10 cm) than Envisat, with no difference between the standard (SGDR) and ALES processing. The lack of valid rmse was observed in both track segments with higher/lower loss of data in land-to-ocean/ocean-to-land transitions, respectively. The analysis of the retracked *Ranges* obtained with AltiKa (SGDR and ALES) showed unrealistic values in the vicinity of land.

TABLE I
Along-Track Mean *RMSE* (in Centimeters) in the Two Track Segments Analyzed (D#0360 and A#0831) With Different Land Masks Applied to the Envisat RA-2 18-Hz Data. The Number of Valid Data Used to Estimate the Mean rmse Is Shown in Parentheses

|  | D#0360 (*E-SoG*) | | A#0831 (*W-SoG*) | |
| --- | --- | --- | --- | --- |
|  | ALES | SGDR | ALES | SGDR |
| No land mask applied | 14.4 (49) | 17.0 (49) | 12.1 (67) | 16.6 (67) |
| Land mask: 1 km | 13.6 (42) | 15.9 (42) | 11.8 (63) | 16.2 (63) |
| Land mask: 2 km | 13.4 (36) | 15.7 (36) | 11.8 (60) | 15.4 (60) |
| Land mask: 3 km | 13.4 (31) | 15.9 (31) | 11.7 (57) | 15.3 (57) |
| Land mask: 4 km | 13.3 (26) | 15.4 (26) | 11.7 (55) | 15.3 (55) |
| Land mask: 5 km | 13.5 (20) | 15.4 (20) | 11.7 (49) | 15.3 (49) |

We estimated the mean value of rmse (Envisat) in the study area, testing the effect of the proximity of land in the calculations. We applied northern and southern land masks of 1 to 5 km from coast before estimating the average of the Envisat rmse along the remainder of the track segment. The lack of available Envisat data in most of *alg-Bay* precluded this analysis. The results are summarized in Table I. ALES gives lower (i.e., better) rmse with little dependence on the land mask extent: values with a land mask of 1 km already approach the asymptotic values with a larger land mask.

## VI. Discussion

The first consideration that needs to be made when discussing the results presented in the previous section is related to the chirp bandwidth. In most of the Envisat cycles for both tracks analyzed, the RA-2 instrument was operating in a nonocean mode when coming out from land and keeping that bandwidth for a few seconds. The overall percentage of nonocean waveforms is higher than seen in other coastal areas probably due to the complex topography, and conversely in a small number of cycles, the chirp bandwidth was found to be 320 MHz even over land: both of these phenomena should be investigated further. In summary, the availability of Envisat data amenable to accurate retracking (i.e., with 320-MHz waveforms) is significantly reduced when the chirp "flag" is taken into account.

The retracking of AltiKa waveforms in the vicinity of land seems to be compromised by the type of transition. Estimates of *Range* using a full-waveform retracker (SGDR) are often wrong especially in land-to-ocean transitions. In some of these cases, even a subwaveform retracker such as ALES is not able to find an estimate of *Range* due to the following: 1) the retracker failing to find a retrackable subwaveform or 2) the subwaveform being too peaky to allow convergence.

The quality of altimeter-derived sea level data in the SoG depends on many factors: instrument, retracking algorithm, data screening, and proximity of the radar measurements of land. AltiKa gives the highest accuracy (in terms of rmse), but the data editing already applied to the SGDR precluded any further assessment of this product close to the coast. The quality of the Envisat RA-2 SLA obtained with the ALES retracker is better than the official product (SGDR) and CTOH. The availability of Envisat data in the vicinity of land depends on the type of ocean/land transition, with more data in ocean-to-land than land-to-ocean transitions, as previously suggested by [62]. The quality of Envisat data degrades in the last 5 km to the coast, regardless of the type of transition; however, along-track rmse averages are robust against the inclusion of points up to 1 km to the coast, especially when the ALES retracker is adopted. The SSB correction, computed for the first time with SWH and $U_{10}$ from ALES (Envisat), improves the quality of the retrieved sea level. This finding reinforces the call for a dedicated sea state bias correction in the coastal zones.

## VII. Summary and Conclusion

In this work, we have analyzed in detail the Envisat altimeter data availability and accuracy in the SoG. SLAs from the official SGDR product and from the ALES retracker were compared against *in situ* tide gauge data located at Tarifa harbor, on the Spanish coast. Other reprocessing schemes (CTOH) and satellites (AltiKa) were also considered in this study.

Data screening in the coastal zone is crucial in order to avoid inaccurate altimeter data. We have followed three criteria for data rejection.

1) Chirp bandwidth (for Envisat only): the switch to the "ocean" bandwidth (320 MHz) in land-to-ocean transitions needed a few seconds in most of the cycles analyzed in the SoG for both track segments. Only the waveforms recorded in ocean mode can be retracked to sufficient precision with the state-of-the-art ocean-oriented retrackers to obtain geophysical information. For this reason, most of the nearshore radar measurements must be rejected in this type of transition. The "ocean" bandwidth is instead kept close to the coast in all of the ocean-to-land transitions of the cycles analyzed. We have concluded that there is a bias to higher data availability for the ocean-to-land versus land-to-ocean transition in case of changes in the chirp bandwidth.
2) Along-track availability and quality of the geophysical corrections: the cycle-by-cycle analysis revealed that all of the corrections presented full availability along the track segments analyzed. This is mainly due to the fact that most of the corrections used are based on models, so no data gaps are expected in the vicinity of the coast. The only exception to this was the sea state bias. This is due to the fact that SSB is linked to the retracking outputs: SWH and $U_{10}$. Any time the estimate of one or both of these two parameters is corrupted, the SSB correction will also be affected. We have demonstrated, however, that SSB recomputation for Envisat using ALES SWH and $U_{10}$ yields a better agreement of the SLA with *in situ* data.

3) Removal of outliers: the rejection of SLA values outside their range of validity demonstrated that the outliers were mainly confined to the coastal strip in both land-to-ocean and ocean-to-land transitions. In the Algeciras Bay, most of the radar measurements were rejected. Two reasons might explain this: 1) the bay is in a land-to-ocean transition, and hence, a number of measurements are excluded due to the instrument operating in a nonocean mode (only for Envisat), and 2) most of the "ocean" waveforms might still contain land or bright target reflections in the footprint area due to the vicinity of land and calm waters to both sides of the track, and this complicates the retrieval of accurate *Ranges*, SWH, and $U_{10}$.

Overall, the results for the reprocessed ALES Envisat are improved compared to the standard (SGDR) and the reprocessed CTOH data sets. The mean along-track rmse in the Strait between ALES and the tide gauge is below 14/12 cm (D#0360/A#0831), which represents about a 20% improvement with respect to the SGDR. The exclusion of nearshore points improved the results slightly (in terms of lower mean along-track rmse), mainly for the SGDR product. AltiKa measurements appear to be the most accurate, showing the lowest rmse against the tide gauge.

For the first time, high-rate SLA data have been derived in the SoG, the confluence of the Atlantic Ocean and the Mediterranean Sea. The validation of the time series of SLA using ground-truth data has demonstrated that a more accurate *SSB* correction improves the comparison against *in situ* data. The availability of data with higher quality will improve the coverage of the coastal zones, especially in challenging areas such as the SoG. This will also increase their use in many applications, such as long-term coastal sea level changes, storm surges, coastal oceanography, etc. The ability to construct longer time series by using both the Envisat and AltiKa missions (although with an unavoidable 2.5-year gap) paves the way to a better characterization of the oceanic processes.

ACKNOWLEDGMENT

The authors would like to thank C. J. González-Mejías for his help on the harmonic analysis of the *in situ* data. The *in situ* data were obtained from the Spanish Institute of Oceanography (IEO). The Envisat RA-2 SGDR products were downloaded from ESA Earthnet Online.

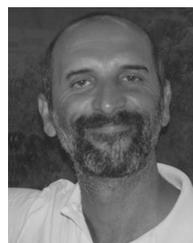

**Jesús Gómez-Enri** received the degree in marine sciences and the Ph.D. degree in satellite oceanography from the University of Cadiz, Cadiz, Spain, in 1996 and 2002, respectively.

From 1999 to 2001, he was a trainee at the European Space Agency European Space Research Institute (ESA-ESRIN), Frascati, Rome, Italy; from 2003 to 2005, he was a Postdoctoral Fellow with the National Oceanography Centre, Southampton, U.K. Since 2005, he has been an Associate Professor with the Applied Physics Department, University of Cadiz. His teaching and research are focused on remote sensing applied to the ocean. He works in satellite radar altimetry and synthetic aperture radar.

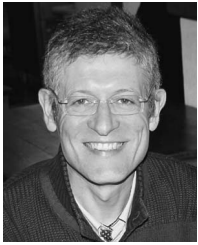

**Paolo Cipollini** (S'93–M'97–SM'03) received the Laurea (M.Eng.) degree in electronics engineering from the University of Pisa, Pisa, Italy, in 1992 and the Ph.D. degree in methods and technologies for environmental monitoring from the University of Florence, Firenze, Italy, in 1996.

He is an Engineer and Satellite Oceanographer with the National Oceanography Centre, Southampton, U.K., working on R&D in satellite radar altimetry and its application to remote sensing of the oceans and the coastal zone. He was the Overall Manager of the ESA development of COASTalt ALTimetry (COASTALT) project (2008–2012) for the development of coastal altimetry for Envisat and is a member of the Ocean Surface Topography Science Team and a Principal Investigator in Cryosat-2 and Sentinel-3 projects.

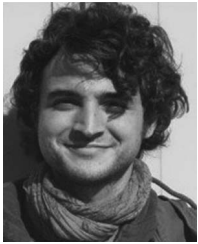

**Marcello Passaro** received the B.S. degree in aerospace engineering from the Politecnico di Milano, Milan, Italy, in 2007, the M.Sc. degree in Earth-oriented space science and technology from the Technische Universitaet Muenchen (TUM), Munich, Germany, in 2009, and the Ph.D. degree from the Graduate School, National Oceanography Centre, Southampton, U.K., in 2016.

He is currently a Research Associate with the German Geodetic Research Institute, TUM. His research is mainly focused on satellite radar altimetry and its application to sea level studies.

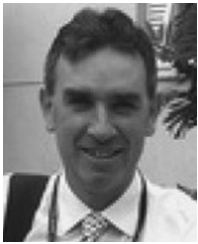

**Stefano Vignudelli** received the Doctoral degree in engineering from the University of Pisa, Pisa, Italy, in 1991.

He is currently a Researcher with the Consiglio Nazionale delle Ricerche (National Research Council), Pisa. He has over 20 years of scientific experience in the area of satellite remote sensing (radar altimetry in particular) for studying coastal and marine/inland environments (water level variability in particular). His major interests include processing methods for data analysis, validation with local field observations, multisensor synergy, and exploitation. His most significant accomplishment has been to lead development of satellite radar altimetry in challenging areas (e.g., coastal zone) to provide improved measurements for water level research and applications.

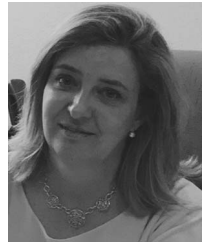

**Begoña Tejedor** received the degree in physics from the Complutense University of Madrid, Madrid, Spain, in 1987, specialist in "Physics of the Earth, Air & Cosmos," and the Ph.D. degree from Las Palmas de Gran Canaria University (LPGCU), Las Palmas de Gran Canaria, Spain, in 1991.

In 1987, she joined the Applied Physics Department (Faculty of Marine Sciences), LPGCU, as an Associate Professor. She joined later on the Faculty of Marine Sciences, University of Cadiz, Cadiz, Spain, as an Associate Professor and the Head of the Applied Physics Department in 1991. Her teaching and research have been focused on physical oceanography. One of her lines of research is the study of physical processes in shallow waters using *in situ* and remote sensing data.

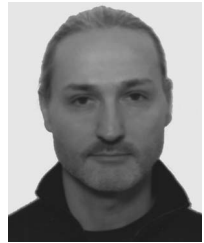

**Josep Coca** received the Bachelor's degree in marine sciences and the Ph.D. degree in marine sciences from the University of Las Palmas de Gran Canaria (ULPGC), Las Palmas de Gran Canaria, Spain, in 1996 and 2006, respectively.

He has developed his work on the applications of satellite oceanography, focused on fisheries, environmental monitoring, and marine ecology. He is also involved in other aspects related to satellite oceanography. He is a Satellite Oceanographer with the SITMA (Integral Marine Technology Service), ULPGC, and a member of the Oceanography and Remote Sensing Group of the Applied Physics Department, University of Cadiz, Cadiz, Spain.